\documentclass[aps,showpacs,showkeys,twocolumn]{revtex4}
\usepackage{mathrsfs}
\usepackage{amssymb}
\usepackage{amsmath}
\usepackage{bm}
\usepackage{graphics}
\usepackage{natbib}
\usepackage{CJK}

\begin{document}

\title{Erasable signature of Majorana bound state due to coupling with the T-shaped quantum-dot structure}

\author{Wei-Jiang Gong$^1$}
\author{Yu-Hang Xue$^1$}
\author{Xiao-Qi Wang$^2$}
\author{Lian-Lian Zhang$^1$}
\author{Guang-Yu Yi$^1$}
\affiliation{1. College of Sciences, Northeastern University, Shenyang 110819, China\\
2. Basic department, Yingkou Institute of Technology, Yingkou 115014, China}
\date{\today}

\begin{abstract}
We theoretically study the transport properties in the T-shaped double-quantum-dot structure, by considering the dot in the main channel to be coupled to the Majorana bound state (MBS) at one end of the topological superconducting nanowire. It is found that the side-coupled dot governs the effect of the MBS on the transport behavior. When its level is consistent with the energy zero point, the MBS contributes little to the conductance spectrum. Otherwise, the linear conductance exhibits notable changes according to the inter-MBS coupling manners. In the absence of inter-MBS coupling, the linear conductance value keeps equal to $e^2\over 2h$ when the level of the side-coupled dot departs from the energy zero point. However, the linear conductance is always analogous to the MBS-absent case once the inter-MBS coupling comes into play. These findings provide new information about the leakage effect of MBSs in quantum-dot structures.
\end{abstract}
 \maketitle

\bigskip

\section{Introduction}
Quantum transport through quantum-dot (QD) systems
has always been one of the main subjects in the field of condensed matter physics, because of the fundamental physics and potential applications of QDs in solid-state physics and quantum computation\cite{QD1,QD2,QD3}.
It is well known that QDs are characterized by the discrete levels and strong Coulomb interactions, which induce abundant quantum transport phenomena, including the well-known resonant tunneling and Kondo resonance\cite{Re1,Re2}. Moreover, multiple QDs can be coupled to form QD molecules with different geometries. Compared with the single QD, QD molecules provide multiple transmission paths for the transport process, and then the quantum interference plays nontrivial roles in adjusting the transport properties\cite{dot1,dot2,dot3,dot4,dot6}. As a result, many interesting results have been observed in the QD-molecule systems, such as the Fano effect\cite{Fano-dot1,Fano-dot2}, Fano-Kondo effect\cite{FKondo}, Aharonov-Bohm effect\cite{A-B-dot}, Dicke effect\cite{Dicke-dot}, and bound states in continuum\cite{BSC-dot}.

\par
Regarding the QD molecules, the double QDs (DQDs) are more typical, especially the T-shaped DQDs. In such systems, the side-coupled QD is important for controlling the transport behaviors. When its level is accordant with the energy zero point of the whole system, the well-defined antiresonance comes into being. This result is attributed to the occurrence of the Fano effect\cite{Fano-dot1}. Just for this reason, the T-shaped DQDs have been proposed as the promising candidate for enhancing the efficiency of thermoelectric effect\cite{Fano-thermoelectric}. It has been reported that in such a
system, the thermoelectric figure of merit $ZT$ can be improved to a great degree by the Fano antiresonance. Besides, by manipulating the spin degree of freedom of the side-coupled QD, the high-efficiency spin polarization can be realized. And then, the T-shaped DQD structure is also a good setup for spintronics\cite{spintronics-T-dot}.
Moreover, some reports have demonstrated that T-shaped DQD geometry is important for observing the two-stage Kondo effect\cite{two-stage Kon-T-dot}.
\par
The successful realization of the Majorana bound states (MBSs) introduces the new connotation to the fundamental physics and applications\cite{exp1,exp2,exp3,exp4,Simon1,Wil,aa1,aa2}. Inspired by their abundant physics, lots of theoretical groups have dedicated themselves to the research in this field\cite{Kitaev1,Sau1,Oreg,Sau2,Sau3,Alice,SatoM}. Various interesting results have been reported. For instance, when a pair of MBSs is coupled to the two leads of one circuit, the nonlocality of the MBSs was observed because of the occurrence of the crossed Andreev reflection\cite{Beenakker,liujie1}. In the junction between a normal metal and a chain of coupled MBSs, the Andreev reflection behavior shows odd-even
effects\cite{Flensberg1,EPL1}. Furthermore, the transport properties of mesoscopic circuits have been investigated by considering finite couplings between the regular bound states and the MBS\cite{Kond1}. It has been demonstrated that the MBS affects the conductance through the noninteracting QD by giving rise to the sharp decrease of the conductance by $1\over2$\cite{Liude,GongMBS,GongCAP}. When the MBSs are indirectly coupled to the leads via QDs, the local and crossed Andreev reflections can be controlled by shifting the QD levels. This realizes the controllable nonlocal transport of MBSs\cite{Liujie2,PRL,aa3}. In addition, MBSs have been found to make nontrivial contributions to the electron correlations\cite{correl1,correl2}.

\par
In view of the properties of the T-shaped DQDs and MBSs, we consider that quantum transport through the T-shaped DQD structure with the side-coupled MBS is certain to display abundant and interesting results. This expectation has obtained the first-step verification\cite{Weymann}. Following this research progress, in the present work we aim to investigate the transport behaviors in the T-shaped DQD system with one MBS coupling laterally to the QD in the main channel. The calculation results show that the side-coupled QD still play its important role in governing the transport property, regardless of the presence of the MBS. To be concrete, when the level of the side-coupled QD is consistent with the energy zero point, the MBS decouples from the T-shaped DQDs and makes zero contribution to the conductance spectrum. Otherwise, the linear conductance exhibits different properties according to the inter-MBS coupling manners. In the absence of inter-MBS coupling, its magnitude keeps equal to $e^2\over 2h$ when the level of the side-coupled QD departs from the energy zero point. However, once the inter-MBS coupling appears, it is always the same as the MBS-absent case.
\par

\begin{figure}
\centering \scalebox{0.43}{\includegraphics{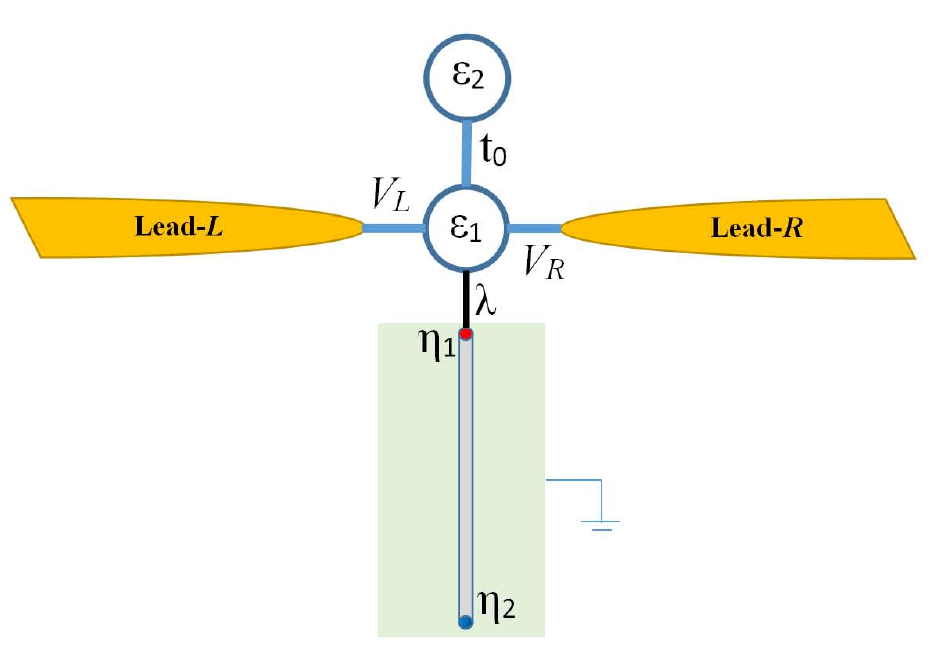}} \caption{
Schematic of a T-shaped DQD structure with the QD in the main channel coupling to MBS-1 (labeled as $\eta_1$). The MBSs are assumed to form at the ends of the one-dimensional topological superconducting nanwire which is achieved by applying magnetic field and superconducting proximity effect to the nanowire with strong spin-orbit interaction. \label{struct}}
\end{figure}
\section{Theoretical model}
Our considered T-shaped DQD structure is illustrated in Fig.\ref{struct}, in which one MBS is supposed to couple to the QD in the main channel. The Hamiltonian of the whole system is written as
$H=H_{C}+H_{DM}+H_{T}$.
$H_C$ is the Hamiltonian of the leads, $H_{DM}$ denotes the Hamiltonian of the QDs, MBSs, as well as their couplings. Each part is given by
\begin{small}
\begin{eqnarray}
H_{C}&=&\sum_{\alpha=L,R;k \sigma}\varepsilon_{\alpha k}c^{\dag}_{\alpha k\sigma}c_{\alpha k\sigma},\\
H_{DM}&=&\sum_{j\sigma}\varepsilon_j d^\dag_{j\sigma}d_{j\sigma}+\sum_{\sigma}(t _0d^\dag_{1\sigma}d_{2\sigma}+h.c.)\notag\\
+\sum_j &U_j& n_{j\uparrow}n_{j\downarrow}+i\epsilon_m\eta_{1}\eta_2+(\lambda d_{1\uparrow}-\lambda^* d^\dag_{1\uparrow})\eta_1.
\label{Hamilton}
\end{eqnarray}
\end{small}
$c^\dag_{\alpha k\sigma}$
($c_{\alpha k\sigma}$) is to create (annihilate) an electron in state $|k\sigma\rangle$ of lead-$\alpha$. $d^\dag_{j\sigma}$ ($d_{j\sigma}$) is the creation (annihilation)
operator for QD-$j$. $\varepsilon_j$ denotes the level of QD-$j$, and $t_0$ is the interdot coupling coefficient. $U_j$ denotes the intradot Coulomb interation strength. Next, $\eta_j$ is the Majorana operator, and $\lambda$ represents the coupling magnitude between QD-1 and MBS-1.
For the expression of $H_T$, it takes the form as
\begin{eqnarray}
H^{}_T&=&\sum_{\alpha k\sigma}\mathcal{V}_{\alpha k}c^\dag_{\alpha k\sigma}d_{1\sigma}+h.c.,
\end{eqnarray}
in which $\mathcal{V}_{\alpha k}$ represents the QD-lead coupling coefficient.
\par
We next proceed to calculate the current passing through our system. The current flow in lead-$\alpha$ can be defined as $J_{\alpha}=-e\langle {\dot{\hat{N}}_{\alpha}}\rangle$ with $\hat{N}_{\alpha}=\sum_{k\sigma}c_{\alpha k\sigma}^\dag c_{\alpha k\sigma}$. Using the Heisenberg equation of
motion, the current can be rewritten as $J_{\alpha}=-e\sum_{k\sigma}[{\cal V}_{\alpha k}G^<_{1\alpha,\sigma}(t,t)+c.c]$, where $G^<_{1\alpha,\sigma}(t,t')=i\langle c^\dag_{\alpha k\sigma}(t')d_{1\sigma}(t)\rangle$ is the lesser Green's function. With the help of the
Langreth continuation theorem and the Fourier
transformation, we have\cite{Formula}
\begin{equation}
J_{\alpha}={e\over h}\int dE \mathrm{Tr}\{{\bf\Gamma}^{\alpha}_e[(G^r-G^a)f_{\alpha e}(E)+G^<]\},
\end{equation}
in which $f_{\alpha e}(E)$ is the electronic Fermi distribution in lead-$\alpha$. $G^{r,a,<}$ are the retarded, advanced, and lesser Green's
functions in the Nambu representation, which are defined as $G^r(t,t')=-i\theta(t-t')\langle\{\Psi(t),\Psi^\dag(t')\}\rangle$
and $G^<(t,t')=i\langle\{\Psi^\dag(t')\Psi(t)\}\rangle$ with $G^a=[G^r]^\dag$.
The field operator is given by $\Psi=[d_{1\uparrow},d^\dag_{1\uparrow},d_{2\uparrow},d^\dag_{2\uparrow},\eta_1,\eta_2; d_{1\downarrow},d^\dag_{1\downarrow},d_{2\downarrow},d^\dag_{2\downarrow}]^T$.
${\bf\Gamma}^{\alpha}_e$ is the linewidth matrix function of
the metallic lead, which describes the coupling strength between the lead and the QDs. If the lead is manufactured by two-dimensional electron gas, the elements of ${\bf\Gamma}^{\alpha}_e$ will be independent of energy.
\par
It is certain that for calculating the current, one must obtain
the expressions of the retarded and lesser Green's functions.
The retarded Green's function can be obtained
from the Dyson's equation. After a straightforward derivation, the retarded Green's function in the noninteracting case can be written out, i.e.,
\begin{small}
\begin{eqnarray}
&&[G^r_\sigma(E)]^{-1}=E \textbf{I}_{\sigma}-H_{DM,\sigma}+{i\over2}{\bf\Gamma}, \label{green}
\end{eqnarray}
\end{small}
where \begin{small}
\begin{eqnarray}
H_{DM,\uparrow}=\left[\begin{array}{cccccc}
\varepsilon_{1}&0&t_0&0&-\lambda^*&0\\
0&-\varepsilon_1&0&-t^*_0&\lambda&0\\
t_{0}&0&\varepsilon_2&0&0&0\\
0&-t_0&0&-\varepsilon_2&0&0\\
-\lambda&\lambda^*&0&0&0&i\epsilon_m\\
0&0&0&0&-i\epsilon_m&0
\end{array}\right]\,  \label{A77}
\end{eqnarray}
\end{small}
and
\begin{small}
\begin{eqnarray}
H_{DM,\downarrow}=\left[\begin{array}{cccc}
\varepsilon_{1}&0&t_0&0\\
0&-\varepsilon_1&0&-t^*_0\\
t_{0}&0&\varepsilon_2&0\\
0&-t_0&0&-\varepsilon_2
\end{array}\right]\,. \label{A77}
\end{eqnarray}
\end{small}
In our system, $[{\bf\Gamma}]_{jl}=\sum_{\alpha}([{\bf\Gamma}_{e}^{\alpha}]_{jl}+[{\bf\Gamma}^{\alpha}_{h}]_{jl})$, and ${\bf\Gamma}^{\alpha}_{e}$ and ${\bf\Gamma}^{\alpha}_{h}$ are respectively defined as
$\Gamma^{\alpha}_{e,jl}=2\pi\delta_{j1}\delta_{l1}\sum_k |\mathcal{V}_{\alpha k}|^2\delta(E-\varepsilon_k)$ and $\Gamma^{\alpha}_{h,jl}=2\pi\delta_{j2}\delta_{l2}\sum_k |\mathcal{V}_{\alpha k}|^2\delta(E+\varepsilon_k)$. Within the wide-band approximation of the lead, we will have $\Gamma^{\alpha}_{e,11}=\Gamma^{\alpha}_{h,22}$. $G^a$ can be solved via the relationship $G^a=[G^r]^\dag$. In this work, we mainly pay attention to the case of left-right symmetry, i.e., $\Gamma^{\alpha}_{e,11}=\Gamma_0$.

As for the lesser Green's function, it can be deduced by using Keldysh equation
$G^<_{\sigma}=G^r_{\sigma}\Sigma^<_{}G^a_{\sigma}$, where
\begin{eqnarray}
\Sigma^<_{}=\left[\begin{array}{ccc} \Sigma^<_{11} &0&\cdots\\
0& \Sigma^<_{22}&\\
\vdots&&\ddots
  \end{array}\right].\
\end{eqnarray}
with $\Sigma^<_{11}=i\Gamma^{L}_{e,11}f_{Le}+i\Gamma^{R}_{e,11}f_{Re}$ and $\Sigma^<_{22}=i\Gamma^{L}_{h,22}f_{Lh}+i\Gamma^{R}_{h,22}f_{Rh}$.
\par
When the intralevel Coulomb interaction is incorporated, the Green's function should be managed within approximations for its solution. In general, the Hubbard-I approximation is feasible to solve the retarded Green's function ${G}^{r}$ if the electron correlation effect is relatively weak\cite{Hubbard}. And then, the only change of the Green's function matrix is mainly manifested as the expression of the QD's part, i.e.,
\begin{small}
\begin{eqnarray}
&&[G^r_{\sigma}(E)]^{-1}=(E \textbf{I}_{\sigma}-H_{DM,\sigma}){\cal R}_{\sigma}+{i\over2}{\bf\Gamma}_{}, \label{greenU}
\end{eqnarray}
\end{small}
where ${\cal R}_{je(h),\sigma}={E\mp\varepsilon_{j}\mp U_{j}\over E\mp\varepsilon_{j}\mp U_{j}\pm U_{j}\langle n_{j\bar{\sigma}}\rangle}$. $\langle n_{j\bar{\sigma}}\rangle$ is the average electron occupation number expressed as $\langle n_{j\sigma}\rangle=-{i\over2\pi}\int d\omega G^{<}_{jj,e\sigma}(E)$.

After the derivation above, the electronic current in the case of left-right symmetry
can be given as
\begin{eqnarray}
J={e\over h}\int T(E)[f_{Le}-f_{Re}]dE,
\end{eqnarray}
in which $T(E)=\sum_{\sigma}T_{\sigma}(E)=-\Gamma_0 \sum_{\sigma}{\rm Im}G^r_{11,\sigma}$ is the transmission function. In the case of zero temperature limit, the current formula can be reexpressed, yielding
$J={e\over h}\int _{-{{eV_b\over2}}}^{{{eV_b\over2}}}T(E)dE$.
It is evident that $T(E)$ is the most critical quantity to evaluate the electronic current. In the noninteracting case, we are allowed to write out the analytical expression of it.

By solving the retarded Green's function matrix in Eq.(\ref{green}), we obtain the expression of $G^r_{11,\sigma}$ and the resulting transmission function in the noninteracting case, i.e.,
\begin{eqnarray}
&&T_{\uparrow}(E)={\Gamma_0^2(E-\varepsilon_2)^2\over |\det [G^r_{\uparrow}]^{-1}|^2}\{(E^2-\epsilon_m^2)^2(E+\varepsilon_2)^2\Gamma_0^2\notag\\
&&+[(E^2-\epsilon_m^2)(E+\varepsilon_1)(E+\varepsilon_2)-(E^2-\epsilon_m^2)t_0^2\notag\\
&&+E(E+\varepsilon_2)\lambda^2]^2\};\label{TW11}\\
&&T_{\downarrow}(E)={\Gamma_0^2(E-\varepsilon_2)^2\over |\det [G^r_{\downarrow}]^{-1}|^2}\{[(E+\varepsilon_1)(E+\varepsilon_2)-t_0^2]^2\notag\\
&&+(E+\varepsilon_2)^2\Gamma^2_0\},
\end{eqnarray}
where $t_0$ and $\lambda$ have been assumed to be real. Following these results, the differential conductance can be discussed because it is defined as
\begin{small}
\begin{equation}
\mathbb{G}_{dif}={\partial J\over\partial V_b}={e^2\over 2h}[T(E={eV_b\over2})+T(E=-{eV_b\over2})].\label{diff}
\end{equation}
\end{small}
Note that at equilibrium, the chemical potential $\mu$ in the metallic leads has been considered to be the energy zero point.

On the other hand, at the zero-bias limit, the current formula can be approximated as $J=\mathcal{G}\cdot V_b$. $\mathcal{G}$, the linear conductance, is also important for describing the transport properties, defined as
\begin{equation}
\mathcal{G}={e^2\over h}T(E=0).\label{linear}
\end{equation}
From Eqs.(11)-(12), the expressions of $T_{\sigma}(E=0)$ is written as
\begin{eqnarray}
&&T_{\uparrow}(E=0)={\Gamma_0^2\varepsilon_2^2[(\varepsilon_1\varepsilon_2-t^2_0)^2+\Gamma_0^2\varepsilon_2^2]\epsilon_m^4\over |\det [G^r_{\uparrow}(E=0)]^{-1}|^2},\notag\\
&&T_{\downarrow}(E=0)={\Gamma_0^2\varepsilon_2^2[(\varepsilon_1\varepsilon_2-t^2_0)^2+\Gamma_0^2\varepsilon_2^2]\over |\det [G^r_{\downarrow}(E=0)]^{-1}|^2}.
\end{eqnarray}
In both cases, the current properties can be clarified by calculating the transmission function.

\section{Numerical results and discussions \label{result2}}
This section proceeds to investigate the transport properties in the T-shaped DQD structure with additional side-coupled MBSs which appear at the ends of the topological superconducting nanowire. In order to present the complete analysis, we would like to concentrate on the differential conductance ${\mathbb{G}}_{dif}$ and the linear conductance $\mathcal{G}$, respectively, since they describe the transport properties from different aspects. For calculation, the level of QD-1 is fixed with $\varepsilon_1=0$, and the temperature of the system is assumed to be zero as well. As for the Coulomb strength, we consider the noninteracting and finite-Coulomb cases, respectively.
\begin{figure}[htb]
\centering \scalebox{0.43}{\includegraphics{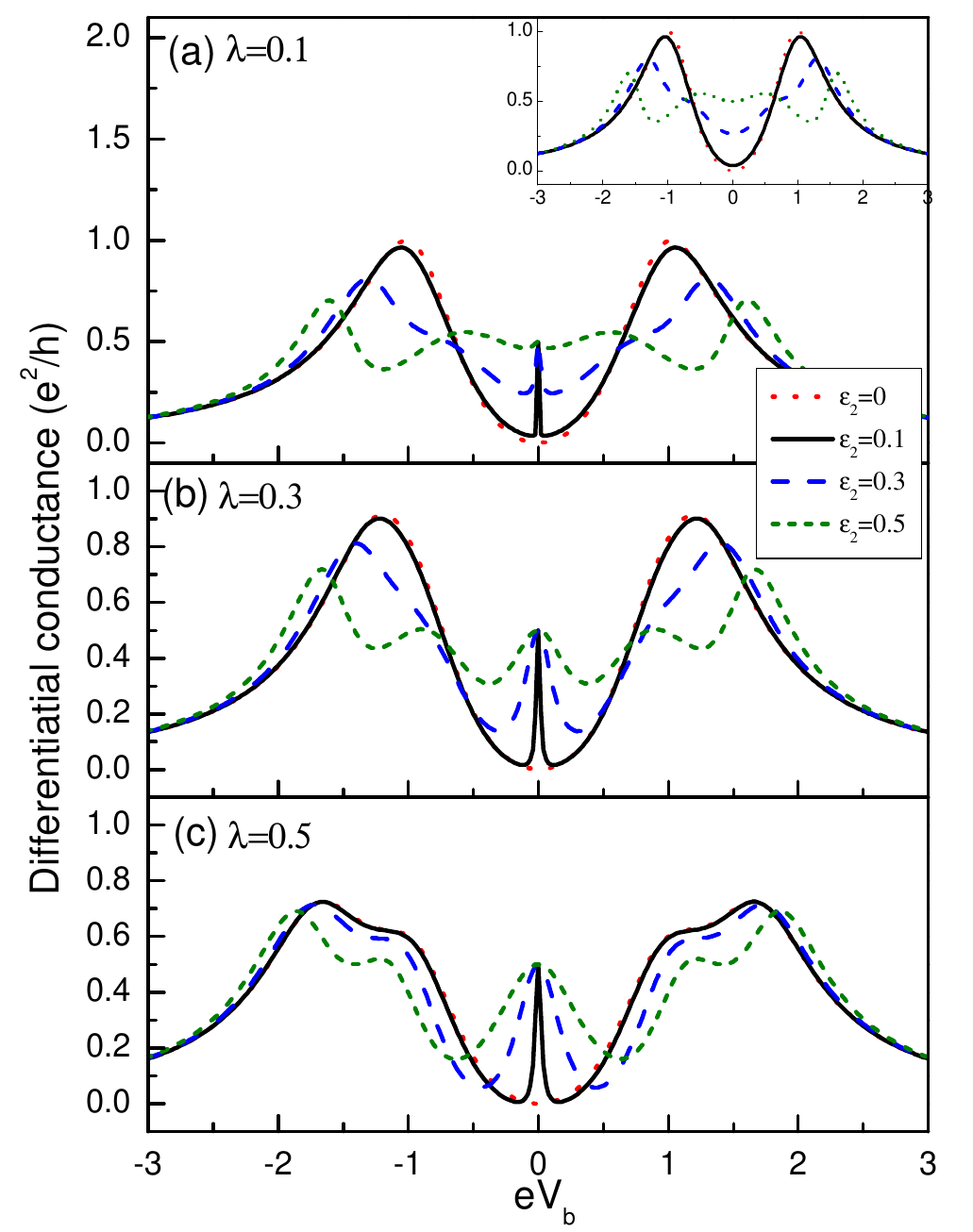}}
\caption{Spin-up component of the differential conductance when the isolated MBS-1 couples to QD-1. (a)-(c) Results of $\varepsilon_2$ from zero to $0.5$ when the MBS-QD coupling coefficient is taken to be $0.1$, 0.3, and 0.5. The insert of (a) shows the conductance of the down-spin component.}
\end{figure}
\par
To describe the basic physics picture of this system, we start with the noninteracting case by taking $\Gamma_0=t_{0}=0.5$. Fig.2 shows the spectra of the spin-up component of the differential conductance in the case of $\epsilon_m=0$, since MBS-1 is assumed to couple to the spin-up states in QD-1. For comparison, the spin-down conductance is also shown in the insert of Fig.2(a). The coupling between MBS-1 and QD-1 is taken to be $\lambda=0.1$, $0.3$, and $0.5$, respectively. In Fig.2(a) where $\varepsilon_2=0$, $0.1$, $0.3$, and $0.5$ with $\lambda=0.1$, we see that at $\varepsilon_2=0$, two peaks exist in the conductance spectrum in the vicinity of $eV_b=\pm1.0$, with one antiresonance point at the zero-bias limit. It is evident that the conductance is identical with the spin-down component, as shown in the insert of Fig.2(a). When the level of QD-2 departs from the energy zero point, e.g., $\varepsilon_2=0.1$, the antiresonance transforms into one peak, and its magnitude is ${1\over2}$ high (in unit of $e^2\over h$). With the increase of $\varepsilon_2$, the whole conductance spectrum is suppressed accordingly, accompanied by the appearance of more conductance peaks. In this process, the zero-bias conductance value is robust, but its corresponding peak is merged following the disappearance of the conductance valley.
Next, Fig.2(b)-(c) show the results of $\lambda=0.3$ and $0.5$, respectively. One can readily find that at $\varepsilon_2=0$, the conductance zero can still be observed at the zero-bias limit. However, the conductance peaks in the high- and low-energy regions are split and suppressed following the increase of $\lambda$. In addition, we see that increasing $\varepsilon_2$ or $\lambda$ induces the effect similar to the widening of the conductance peaks at the zero-bias limit. These results indicate that when an isolated MBS is coupled to the QD in the main channel of the T-shaped DQD circuit, it contributes to the quantum transport in different ways when the level of the side-coupled QD coincides with or departs from the energy zero point. To be concrete, when the level of this QD is fixed at the energy zero point, the low-bias conductance spectra are analogous with the MBS-absent result. Otherwise, if $\varepsilon_2$ is not equal to zero, the influence of the MBS will become apparent, i.e., manifested as the existence of halved zero-bias peak.
\par
From of Eq.(11) and Eq.(13), the results in Fig.2 can be clarified. For $\varepsilon_1=\epsilon_m=0$,
\begin{small}
\begin{eqnarray}
&&T_{\uparrow}(E)={\Gamma_0^2(E-\varepsilon_2)^2\over {\cal D}}[E^2t_0^4-2t^2_0E(E+\varepsilon_2)(E^2-\lambda^2)\notag\\
&&+(E+\varepsilon_2)^2(E^4+2\lambda^4+E^2\Gamma_0^2-2E^2\lambda^2)],\label{TW11n}
\end{eqnarray}
\end{small}
with ${\cal D}=E^2t_0^8-4t_0^2(E^2-\varepsilon^2_2)E^2(E^2-\lambda^2)(E^2+\Gamma_0^2-2\lambda^2)
-4t_0^6E^2(E^2-\lambda^2)+(E^2-\varepsilon^2_2)^2(E^2+\Gamma_0^2)[(E^2-2\lambda^2)^2+E^2\Gamma_0^2]+2t_0^4E^2[3E^4+2\lambda^4+E^2(\Gamma_0^2-6\lambda^2)-\varepsilon_2^2(E^2-\Gamma_0^2-2\lambda^2)]$.
In the case of $\varepsilon_2=0$,
$T_{\uparrow}(E)={1\over2}[{E^2\Gamma_0^2 \over (E^2-t_0^2)^2+E^2\Gamma_0^2}+{E^2\Gamma_0^2\over (E^2-t_0^2-2\lambda^2)^2+E^2\Gamma_0^2}]$.
In the limit of $E\to0$, the transmission is weakened to be zero completely. And then, the antiresonance is robust and independent of the structural parameters. In addition, the above equation helps us to understand the four-peak structure of the conductance spectrum in this case. When $E=\pm t_0$ or $\pm \sqrt{t_0^2+2\lambda^2}$, $T_{\uparrow}(E)$ will reach its maximum. In fact, we find from Eq.(6) that in the case of $\epsilon_m=\varepsilon_2=0$, the QD molecule should possess five eigenlevels. Except the four above, another level is located at the energy zero point. This can be verified by solving the Hamiltonian $H_{DM,\uparrow}$ as follows. For $\epsilon_m=0$, $H_{DM,\uparrow}$ is simplified to be five-dimensional matrix, i.e,
\begin{eqnarray}
H_{DM,\uparrow}=\left[\begin{array}{ccccc}
\varepsilon_{1}&0&t_0&0&-\lambda^*\\
0&-\varepsilon_1&0&-t^*_0&\lambda\\
t_{0}&0&\varepsilon_2&0&0\\
0&-t_0&0&-\varepsilon_2&0\\
-\lambda&\lambda^*&0&0&0
\end{array}\right]\,.  \label{A77}
\end{eqnarray}
The eigenvalues are
$E_{1}=0$ and $E_{n\neq1}={\pm1\over\sqrt{2}}\sqrt{\varepsilon_1^2+\varepsilon_2^2+2(t^2_0+\lambda^2)\pm\sqrt{\Delta}}$ with $\Delta=(\varepsilon_1+\varepsilon_2)^2[(\varepsilon_1-\varepsilon_2)^2+4t^2_0]+4(\varepsilon_1^2-\varepsilon_2^2)\lambda^2+4\lambda^4$.
It shows that one zero-energy state exists in this case, independent of the tuning of the structural parameters. It is surely the antiresonance effect that eliminates its corresponding conductance peak. Alternatively when the level of QD-2 departs from its zero value, the antiresonance vanishes and the zero-bias conductance peak comes into being. One can readily find in Eq.(14) that in the case of $E\to0$, $T_{\uparrow}(E)={2\Gamma_0^2\varepsilon_2^4\lambda^4\over4\Gamma_0^2\varepsilon_2^4\lambda^4}={1\over2}$. Therefore, once the level of the side-coupled QD tunes away from the energy zero point, one peak is certain to arise at the zero-energy limit with its magnitude being $1\over2$ when an isolated MBS is coupled laterally to the QD in the main channel.

\par
\begin{figure}[htb]
\centering \scalebox{0.43}{\includegraphics{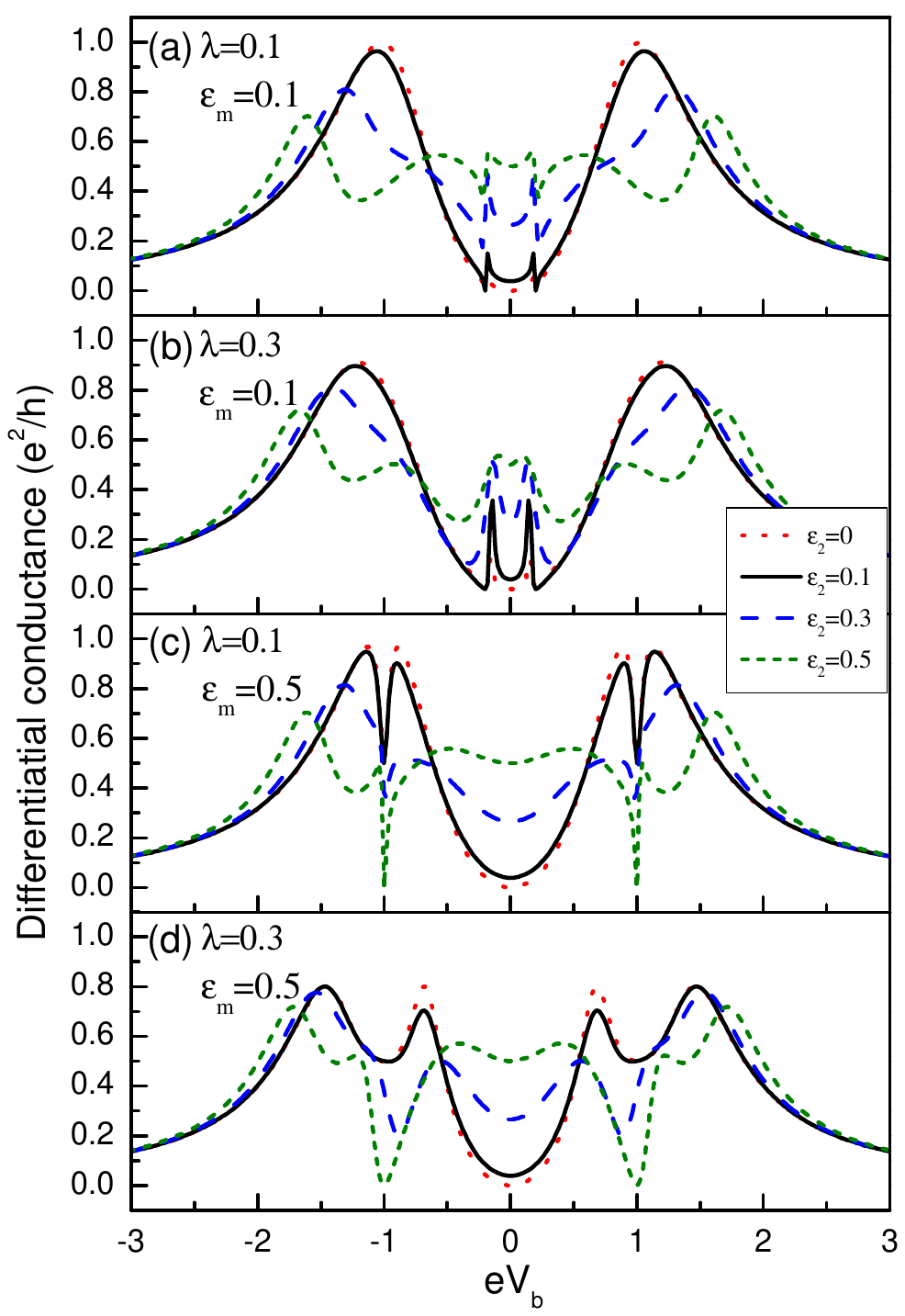}} \caption{Spin-up component of the differential conductance in the cases of $\epsilon_m=0.1$ and 0.5, respectively. (a)-(b) Results of $\epsilon_m=0.1$ with $\lambda=0.1$ and $0.3$. (c)-(d) Conductance of $\epsilon_m=0.5$.
  \label{struct2}}
\end{figure}
\par

\par
To uncover the conductance transition caused by the change of the nonlocality of the MBSs, we would like to evaluate the conductance properties in the presence of inter-MBS coupling. The numerical results are shown in Fig.3. Firstly, the results of weak inter-MBS coupling, e.g., $\epsilon_m=0.1$, are shown in Fig.3(a)-(b). The coupling between QD-1 and MBS-1 is taken to be $\lambda=0.1$ and $0.3$, respectively. It can be clearly found that the coupled MBSs contribute to the quantum transport in an alternative way. Compared with the results in Fig.2(a)-(b), the leading conductance spectra are similar to the case of $\epsilon_m=0$. The notable change is that the zero-bias conductance peak splits into two in the presence of inter-MBS coupling. Accordingly, two peaks appear on the two sides of $eV_b=0$, the distance of which is related to the inter-MBS coupling. Note that in this case, the peak heights are proportional to the QD-MBS coupling, especially for $\varepsilon_2$ close to zero. As for the role of QD-2, it shows that with the increase of $\varepsilon_2$, the conductance peaks in the low-bias region are enhanced and widened gradually. This result can be explained as follows. When the level of QD-2 departs away from the energy zero point, the destructive effect of the transport process becomes weak, and then the MBSs play the dominant role.
Fig.3(c)-(d) show the conductance spectra of strong inter-MBS coupling, i.e., $\epsilon_m=0.5$. It can be seen that the role of QD-2 is accordant with the case of $\epsilon_m=0.1$, but the conductance peaks vary in the other way. As shown in Fig.3(c) where $\lambda=0.1$, the conductance dips appear around $eV_b=\pm1.0$ with the increase of $\varepsilon_2$. And when $\varepsilon_2=0.5$, the conductance dip changes to be the antiresonance point. Next, when the QD-MBS coupling increases, e.g., $\lambda=0.3$, similar results can be found. The difference is that increasing the QD-MBS coupling causes a more apparent peak-to-dip phenomenon. At $\varepsilon_2=0.5$, the antiresonance also has the opportunity to occur with the wider antiresonance valley.
\begin{figure}[htb]
\centering \scalebox{0.53}{\includegraphics{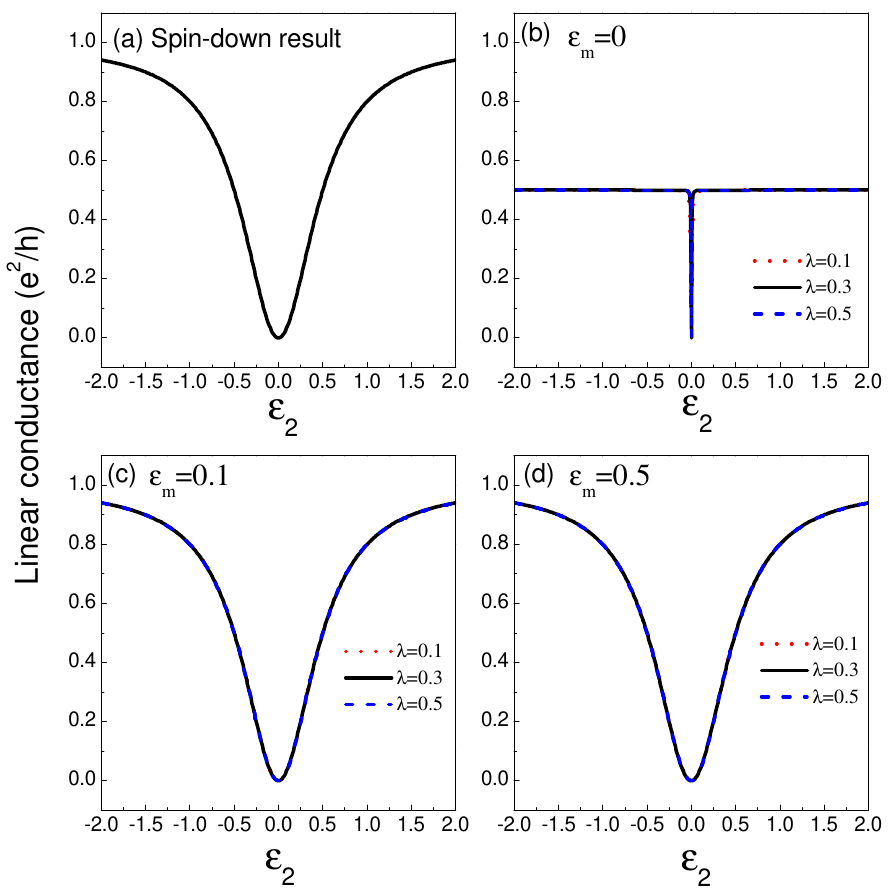}} \caption{Linear-conductance curves in different cases. (a) Spin-down result of the linear conductance. (b)-(d) Spin-up conductance in the cases of $\epsilon_m=0$, $0.1$, and $0.5$.
  \label{struct3}}
\end{figure}
\par
We next analyze the above result with Eq.(11). It also shows that in the presence of the inter-MBS coupling, the transmission function of $\varepsilon_2=0$ is given as
\begin{equation}
T_{\uparrow}(E)={E^2\Gamma_0^2(\mathcal{A}+\mathcal{B})\over[(E^2-t_0^2)^2+E^2\Gamma_0^2](\mathcal{A}+2\mathcal{B})}
\end{equation}
with $\mathcal{A}=(E^2-\epsilon_m^2)^2[(E^2-t_0^2)^2+E^2\Gamma_0^2+(E+\varepsilon_1)^2E^2-2(E+\varepsilon_1)Et_0^2]$ and $\mathcal{B}=E^2\lambda^2\{E^2\lambda^2+2(E^2-\epsilon_m^2)[E(E+\varepsilon_1)-t_0^2]\}$. Based on this result, the zero-bias antiresonance can be well understood. On the other hand, one can see that in the case of $\varepsilon_1=0$,
\begin{eqnarray}
&&T_{\uparrow}(E)={\Gamma_0^2(E-\varepsilon_2)^2\over |\det [G^r_{\uparrow}]^{-1}|^2}\{(E^2-\epsilon_m^2)^2(E+\varepsilon_2)^2\Gamma_0^2\notag\\
&&+[(E^2-\epsilon_m^2)(E^2+E\varepsilon_2-t_0^2)+(E^2+E\varepsilon_2)\lambda^2]^2\}.\notag\\ \label{TW11}
\end{eqnarray}
This exactly means that under the condition of $\varepsilon_2=\epsilon_m$, the antiresonance is allowed to occur at the position of $E=\pm\epsilon_m$. We then understand the double-antiresonance phenomenon co-influenced by the side-coupled QD and coupled MBSs.

\par

Considering the differential conductance properties modified by the MBS, we would like to plot the linear conductance spectra by taking $eV_b\to 0$. The numerical results are shown in Fig.4. Here we also present the spin-down result for comparison [see Fig.4(a)], which has been well-known during the past years. In Fig.4(b) it can be found that in the presence of $\epsilon_m=0$, the conductance plateau encounters its dip in the critical case of $\varepsilon_2=0$, whereas its value remains at $1\over2$ throughout the energy region. As for the change of QD-MBS coupling, it plays a trivial role in modifying the conductance spectrum. Such a result can be proved from the expression of $T_{\uparrow}(E)$ in Eq.(16). Instead, in the case of $\epsilon_m\neq0$, the conductance profile becomes more dependent on the shift of the level of QD-2, as shown in Fig.4(c)-(d). What is notable is that the conductance value does equal to zero in the case of $\varepsilon_2=0$. And moreover, the conductance results are independent of the changes of QD-MBS and inter-MBS couplings. Following a straightforward derivation, we get the expression of $T_{\uparrow}(E=0)$ in the case of $\epsilon_m\neq0$ with $E=0$,
\begin{equation}
T_{\uparrow}(E=0)={\varepsilon_2^2\Gamma_0^2\over (t_0^2-\varepsilon_1\varepsilon_2)^2+\varepsilon_2^2\Gamma_0^2}.
\end{equation}
This result is irrelevant to the parameters $\epsilon_M$ and $\lambda$. Meanwhile, it is the same as that of $T_{\downarrow}(E=0)$. Therefore, for this case, the MBSs become decoupled from the QD in the main channel in the zero-bias limit. Up to now, we find that in this T-shaped DQD structure, the role of the MBSs is tightly dependent on the side-coupled QD. As a typical case with $\varepsilon_2=0$, the signature of the MBSs can be suppressed completely.
\begin{figure}
\centering \scalebox{0.45}{\includegraphics{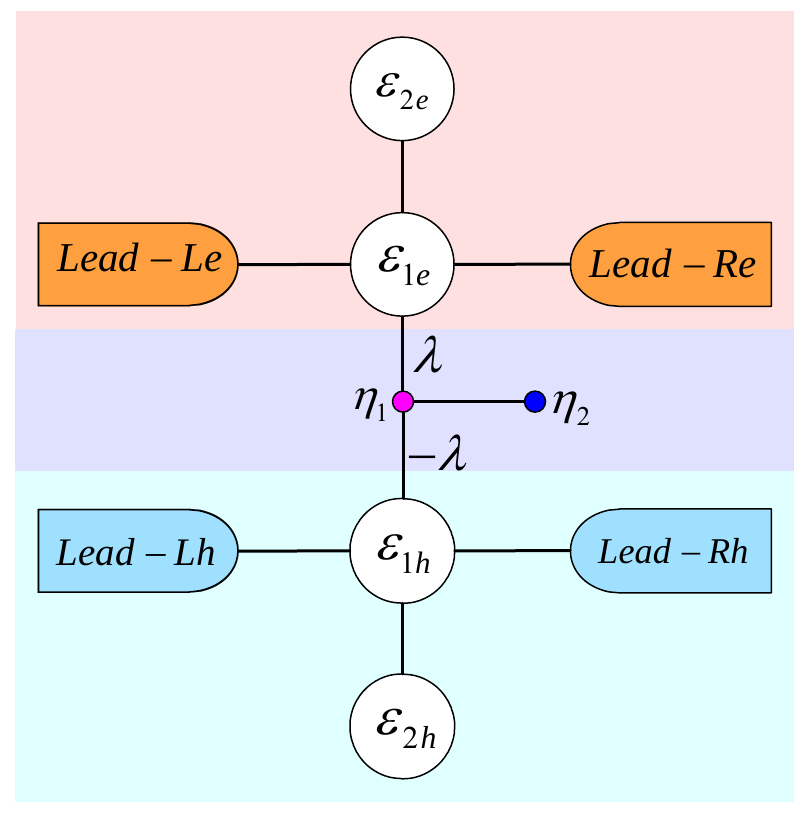}} \caption{Spin-up illustration of our considered T-shaped DQDs in the Nambu representation. The electron and hole parts of the QDs and the MBS part are colored differently for comparison.
  \label{struct4}}
\end{figure}
\par
Next, we present an explanation about the MBS-assisted transport results. To begin with, we plot the schematic of this system in the Nambu representation, as shown in Fig.5. It is not difficult to find that in this representation, our considered structure is just transformed into the geometry of three T-shaped parts coupling serially, i.e., the electronic and hole parts of the DQDs and the MBS part, respectively. Therefore, the characteristic of the T-shaped meso-structure certainly plays its role in governing the transport behaviors. Namely, the side-coupled part induces the destructive quantum interference. We then perform discussions following this idea. The first step is to write out the expression of the retarded Green's function, i.e.,
\begin{eqnarray}
G^r_{11,\uparrow}={1\over E-\varepsilon_1+i\Gamma_0-{t_0^2\over E-\varepsilon_2+i0^+}-\lambda^2 G^r_{m1}},
\end{eqnarray}
where $G^r_{m1}$ is the retarded Green's function of MBS-1, defined as
\begin{eqnarray}
G^r_{m1}={1\over E+i0^+-{\epsilon_m^2\over E+i0^+}-{\lambda^2\over E+\varepsilon_1+i\Gamma_0-{t_0^2\over E+\varepsilon_2+i0^+}}}.
\end{eqnarray}
In Eq.(21), it can be found that the role of QD-2 is indeed dominant. In the case of $E=\varepsilon_2$, $G^r_{11,\uparrow}$ will get close to zero and the transmission is forbidden, independent of the presence of MBSs. The underlying reason should be attributed to the completely destructive interference effect induced by the side-coupled QD. Such a result is surely helpful in clarifying the results of the differential conductance. In the case of $\varepsilon_2=0$, the zero-bias peak is eliminated. Also, note that under the condition of $E=-\varepsilon_2$, the hole state of QD-1 will decouple from MBS-1 since its corresponding Green's function $G^r_{22,\uparrow}\to 0$. In this case, the destructive interference effect of MBS-2 is clearly observed. And if $\epsilon_m=\varepsilon_2$, $G^r_{m1}$ will be equal to infinity. This means that the MBSs contribute to the destructive interference effect during the electron transmission process. Therefore, we can understand the results in Fig.3(c)-(d). On the other hand, when focusing on the result of $E=0$, we see that if $\epsilon_m\neq0$, there will be $G^r_{m1}=0$. And then, the effect of the MBSs disappears, irrelevant to the change of the QD-MBS and inter-MBS couplings. For $\epsilon_m=0$, $G^r_{m1}(E\to0)\approx -[E+\varepsilon_1+i\Gamma_0-{t_0^2\over E+\varepsilon_2+i0^+}]/\lambda^2$, and $G^r_{11,\uparrow}(E\to0)\approx 1/2[i\Gamma_0-{t_0^2E\over E^2-\varepsilon^2_2}]$. Surely, once the level of QD-2 departs from the energy zero point, its impact on $G^r_{11,\uparrow}(E\to0)$ will be erased completely. All these results contribute to the understanding of the linear conductance properties. Based on the above analysis, we can also anticipate that even finite coupling occurs between MBS-1 and QD-2, the above picture also exists at the low-energy limit. Take the case of $\epsilon_m\neq0$ as an example, the addition coupling between MBS-1 and QD-2 cannot change the antiresonance effect induced by MBS-2. This help us to further understand the transport picture in this T-shaped DQD structure.
\par

\begin{figure}[htb]
\centering \scalebox{0.43}{\includegraphics{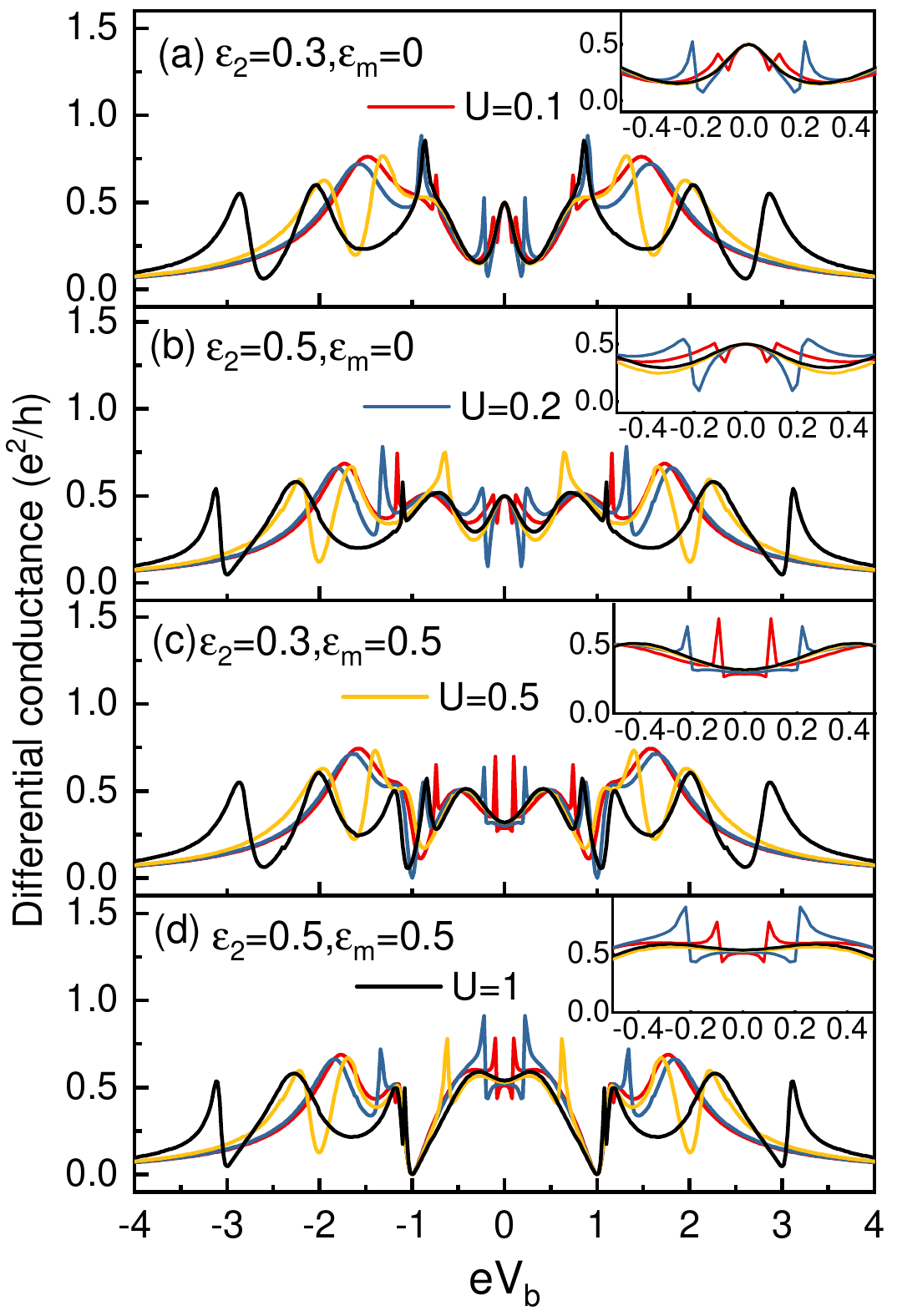}} \caption{Coulomb effect on the differential conductances of the T-shaped DQDs with coupled MBSs. The Coulomb strength takes to be 0.1, 0.2, 0.5, and 1.0, respectively, and the QD-MBS coupling is fixed at $\lambda=0.3$. (a)-(b) shows the results of $\varepsilon_2=0.3$ and $0.5$ with, in the case of $\epsilon_m=0$. (c)-(d) Corresponding results of $\epsilon_m=0.5$.  \label{struct4}}
\end{figure}
Following the noninteracting results, we next incorporate the intradot Coulomb interaction to present the modification of the noninteracting results. If the electron correlation effect is relatively weak, the Hubbard-I approximation is feasible to deal with the Coulomb terms in the Hamiltonian for solving the Green's functions. It is known that the leading effect of the Hubbard-I approximation is to induce the level splitting of the QDs, i.e., from $\varepsilon_j$ to $\varepsilon_j$ and $\varepsilon_j+U_j$. Thus, it can be anticipated that in this system, the quantum transport results will be complicated by the Coulomb repulsions in the QDs. In Fig.6, we plot the differential conductance spectra modified by the electron interaction, where the Coulomb strengths in the QDs are supposed to be the same, i.e., $U_j=U$. For the structural parameters, we take $\varepsilon_1=0$ and $\lambda=0.3$. The results of $\epsilon_m=0$ are shown in Fig.6(a)-(b), where $\varepsilon_2=0.3$ and $0.5$, respectively. It is found that for the weak Coulomb strength, e.g., $U\le0.2$, the conductance peak of the zero-bias limit splits into three. However, the central peak is still localized at the zero-bias limit and its magnitude remains at $1\over2$. When the Coulomb interaction is further enhanced, the conductance peaks increase and move in the repulsive way. And then, the zero-bias peak becomes clear again with the invariant magnitude. Next in the presence of inter-MBS coupling, the change manner of the conductance peaks is similar to the case of $\epsilon_m=0$, as shown in Fig.6(c)-(d). The notable result is that regardless of the increase of Coulomb interaction, the antiresonance positions in the conductance spectrum are robust in the situation of $\varepsilon_2=\epsilon_m$.

\begin{figure}[htb]
\centering \scalebox{0.34}{\includegraphics{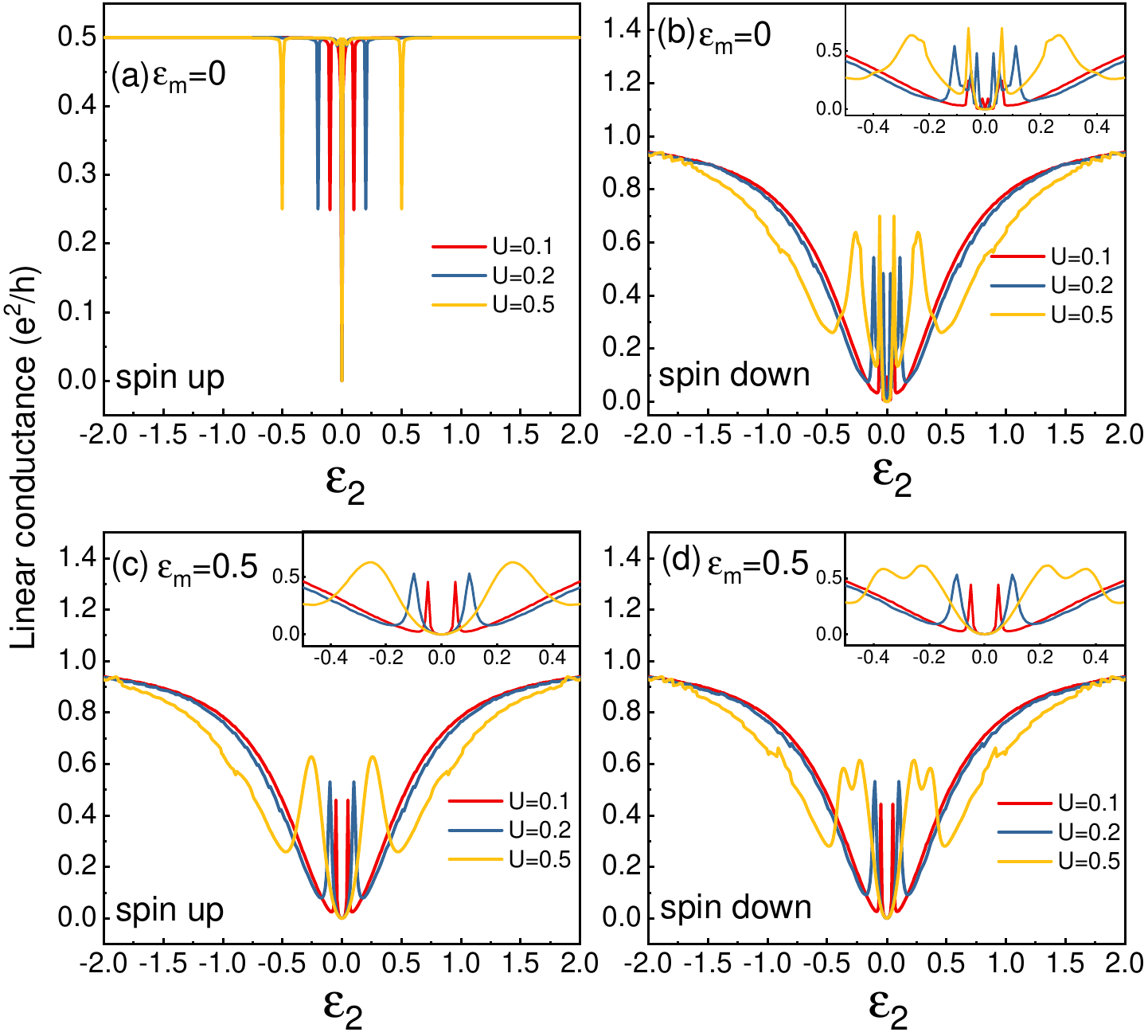}} \caption{Spectra of linear conductances in the presence of intradot Coulomb interaction. Relevant parameters are taken to be $\varepsilon_1=0$ and $\lambda=0.3$. (a)-(b) Spin up and down results for $\epsilon_m=0$. (c)-(d) Conductances under the condition of $\epsilon_m=0.5$.  \label{struct4}}
\end{figure}
\par
In Fig.7, we plot the linear conductance curves when the intradot Coulomb interactions are incorporated. The uniform Coulomb strength in the QDs is taken to be 0.1, 0.2, and 0.5, respectively. For $\epsilon_m=0$, Fig.7(a) shows that the electron interactions in the QDs induce new dips in the conductance spectrum at $\varepsilon_2=\pm U$. This surely arises from the level splitting caused by the Coulomb repulsion within the Hubbard-I approximation. In addition, we readily see that the leading result in the noninteracting case remains, since the conductance magnitude keeps equal to $1\over2$ throughout the energy region. On the other hand, the spin-down conductance manifests as the alternative result, as shown in Fig.7(b). The conductance magnitude varies with the change of $\varepsilon_2$, and at $\varepsilon_2=0$ the antiresonance phenomenon occurs. In such a case, the Coulomb interaction leads to the appearance of the subpeak near the antiresonance point, the distance of which depends on the Coulomb strength. Next, Fig.7(c)-(d) show the results of $\epsilon_m=0.5$. We find that the opposite-spin components of the linear conductance are similar to each other, especially in the weak-Coulomb limit. Thus, despite the Coulomb interaction, the MBS tends to decouple from T-shaped DQDs. Next, the difference between the opposite-spin conductances begins to appear gradually, with the enhancement of Coulomb interaction. And they exhibit different oscillations when $U=0.5$. The underlying reason should be attributed to the different spin occupations in the QDs due to the coupling of the MBSs to the spin-up state. Up to now, we have known the transport properties in this structure in the case of weak Coulomb interaction. That is in the weak-correlation regime, the interplay between the side-coupled QD and the MBS is basically accordant with the noninteracting case in modifying the differential conductance properties.

\section{summary\label{summary}}
In summary, we have performed studies about the transport properties in the T-shaped DQD structure, by introducing one MBS to couple to the QD in the main channel. As a result, it has been found that the influence of the MBS is tightly determined by the level of side-coupled QD. One of typical results is that when the side-coupled QD level is tuned to the energy zero point, the MBS tends to decouple from the DQDs since it contributes little to the conductance spectrum. Otherwise, the linear conductance exhibits clear changes according to the inter-MBS coupling manners. In the case of zero inter-MBS coupling, the linear conductance value keeps equal to $e^2\over 2h$ when the level of the side-coupled QD is away from the energy zero point. Nevertheless, the linear conductance is always identical with the MBS-absent case once the inter-MBS coupling takes place. Therefore, different from the other QD systems, the leakage effect of the MBS depends on the side-coupled QD particularly. This work provides new content for describing the interplay between the QD and MBSs in mesoscopic systems.
Also, we would like to discuss the experimental realization and measurement of our structure. In comparison with the structure in Ref.\cite{aaa}, our system can be achieved by applying more gate voltages to form the second QD and introducing two leads to coupled with the QD neighboring the MBS. By measuring the conductance between the leads, our obtained results can be checked. Therefore, the leading results in this work can be realized and measured according to the nowaday experimental conditions.

\section*{Acknowledgments}
\par
This work was financially supported by the LiaoNing Revitalization Talents Program (Grant No. XLYC1907033), the Fundamental Research Funds for the Central Universities (Grant No. N2002005), and the
National Natural Science Foundation of China (Grant No.
11905027).



\clearpage

\bigskip

\end{document}